\documentclass[10pt]{article}

\usepackage{amsmath}
\usepackage{amssymb}

\usepackage{graphicx}

\usepackage{cite}

\usepackage{color} 

\usepackage[labelfont=bf,labelsep=period,justification=raggedright]{caption}

\makeatletter
\renewcommand{\@biblabel}[1]{\quad#1.}
\makeatother

\date{}

\begin{document}

\begin{flushleft}
{\Large
\textbf{A Reference-Free Algorithm for Computational Normalization of Shotgun Sequencing Data}
}
\\
C. Titus Brown$^{1,2,\ast}$,
Adina Howe$^{2}$,
Qingpeng Zhang$^{1}$,
Alexis B. Pyrkosz$^{3}$,
Timothy H. Brom$^{1}$
\\
\bf{1} Computer Science and Engineering, Michigan State University,
East Lansing, MI, USA
\\
\bf{2} Microbiology and Molecular Genetics, Michigan State University,
East Lansing, MI, USA
\\
\bf{3} USDA Avian Disease and Oncology Laboratory, East Lansing, MI, USA
\\
$\ast$ E-mail: ctb@msu.edu
\end{flushleft}

\section*{Abstract}

Deep shotgun sequencing and analysis of genomes, transcriptomes,
amplified single-cell genomes, and metagenomes has enabled
investigation of a wide range of organisms and ecosystems.  However,
sampling variation in short-read data sets and high sequencing
error rates of modern sequencers present many new computational
challenges in data interpretation.  These challenges have led to the
development of new classes of mapping tools and {\em de novo}
assemblers.  These algorithms are challenged by the
continued improvement in sequencing throughput.  We here describe
digital normalization, a single-pass computational algorithm that
systematizes coverage in shotgun sequencing data sets, thereby
decreasing sampling variation, discarding redundant data, and removing
the majority of errors.  Digital normalization substantially reduces
the size of shotgun data sets and decreases the memory and time
requirements for {\em de novo} sequence assembly, all without
significantly impacting content of the generated contigs.  We apply
digital normalization to the assembly of microbial genomic data,
amplified single-cell genomic data, and transcriptomic data.  Our
implementation is freely available for use and modification.

\section*{Author Summary}

\section*{Introduction}

The ongoing improvements in DNA sequencing technologies have led to a
new problem: how do we analyze the resulting large sequence data sets
quickly and efficiently? These data sets contain millions to billions
of short reads with high error rates and substantial sampling
biases \cite{pubmed19997069}.  The vast quantities of deep sequencing
data produced by these new sequencing technologies are driving
computational biology to extend and adapt previous approaches to
sequence analysis.  In particular, the widespread use of deep shotgun
sequencing on previously unsequenced genomes, transcriptomes, and
metagenomes, has resulted in the development of several new approaches
to {\em de novo} sequence assembly \cite{pubmed20211242}.

There are two basic challenges in analyzing short-read sequences from
shotgun sequencing. First, deep sequencing is needed for complete
sampling. This is because shotgun sequencing samples randomly from a
population of molecules; this sampling is biased by sample content and
sample preparation, requiring even deeper sequencing. A human genome
may require 100x coverage or more for near-complete sampling, leading
to shotgun data sets 300 GB or larger in size\cite{pubmed21187386}.
Since the lowest abundance molecule determines the depth of coverage
required for complete sampling, transcriptomes and metagenomes
containing rare population elements can also require similarly
deep sequencing.

The second challenge to analyzing short-read shotgun sequencing is the
high error rate.  For example, the Illumina GAII sequencer has a 1-2\% error
rate, yielding an average of one base error in every 100 bp of data
\cite{pubmed19997069}.  The total number of errors grows linearly with
the amount of data generated, so these errors usually dominate
novelty in large data sets \cite{pubmed21245053}.  Tracking this
novelty and resolving errors is computationally expensive.

These large data sets and high error rates combine to provide a third
challenge: it is now straightforward to generate data sets that cannot
easily be analyzed \cite{pubmed21867570}.  While hardware approaches
to scaling existing algorithms are emerging, sequencing capacity
continues to grow faster than computational capacity
\cite{pubmed20441614}.  Therefore, new algorithmic approaches to
analysis are needed.

Many new algorithms and tools have been developed to tackle large and
error-prone short-read shotgun data sets. A new class of alignment
tools, most relying on the Burrows-Wheeler transform, has been created
specifically to do ultra-fast short-read alignment to reference
sequence \cite{pubmed19430453}.  In cases where a reference sequence
does not exist and must be assembled {\em de novo} from the sequence
data, a number of new assemblers have been written, including ABySS,
Velvet, SOAPdenovo, ALLPATHS, SGA, and Cortex
\cite{pubmed19251739,pubmed18349386,pubmed20511140,pubmed21187386,pubmed22156294,cortex}.
These assemblers rely on theoretical advances to store and assemble
large amounts of data \cite{pubmed22068540,pubmed20529929}.  As
short-read sequencing has been applied to single cell genomes,
transcriptomes, and metagenomes, yet another generation of assemblers
has emerged to handle reads from abundance-skewed populations of
molecules; these tools, including Trinity, Oases, MetaVelvet,
Meta-IDBA, and Velvet-SC, adopt local models of sequence coverage to
help build assemblies
\cite{pubmed21572440,pubmed22368243,metavelvet,pubmed21685107,pubmed21926975}.
In addition, several ad hoc strategies have also been applied to reduce
variation in sequence content from whole-genome amplification
\cite{pubmed19724646,pubmed22028825}.
Because these tools all rely on k-mer approaches and require exact
matches to construct overlaps between sequences, their performance is
very sensitive to the number of errors present in the underlying data.
This sensitivity to errors has led to the development of a number of
error removal and correction approaches that preprocess data prior to
assembly or mapping
\cite{pubmed21685062,pubmed15059830,pubmed21114842}.


Below, we introduce ``digital normalization'', a single-pass algorithm
for elimination of redundant reads in data sets.  Critically, no
reference sequence is needed to apply digital normalization.  Digital
normalization is inspired by experimental normalization techniques
developed for cDNA library preparation, in which hybridization
kinetics are exploited to reduce the copy number of abundant
transcripts prior to sequencing\cite{pubmed8889548,pubmed7937745}.
{\em Digital} normalization works after sequencing data has been
generated, progressively removing high-coverage reads from shotgun
data sets.  This normalizes average coverage to a specified value,
reducing sampling variation while removing reads, and also removing
the many errors contained {\em within} those reads.  This data and
error reduction results in dramatically decreased computational
requirements for {\em de novo} assembly.  Moreover, unlike experimental
normalization where abundance information is removed prior to sequencing,
in digital normalization this information can be recovered from the
unnormalized reads.




We present here a fixed-memory implementation of digital normalization
that operates in time linear with the size of the input data.  We then
demonstrate its effectiveness for reducing compute requirements for
{\em de novo} assembly on several real data sets.  These data sets
include {\em E. coli} genomic data, data from two single-cell
MD-amplified microbial genomes, and yeast and mouse mRNAseq.

\section*{Results}

\subsection*{Estimating sequencing depth without a reference assembly}

Short-read assembly requires deep sequencing to systematically sample
the source genome, because shotgun sequencing is subject to both
random sampling variation and systematic sequencing biases.  For
example, 100x sampling of a human genome is required for recovery of
90\% or more of the genome in contigs $>$ 1kb \cite{pubmed21187386}.
In principle much of this high-coverage data is redundant and could be
eliminated without consequence to the final assembly, but determining
which reads to eliminate requires a per-read estimate of coverage.
Traditional approaches estimate coverage by mapping reads to an
assembly.  This presents a chicken-and-egg problem: to
determine which regions are oversampled, we must already have an
assembly!

\begin{figure}[!ht]
\centerline{\includegraphics[width=4in]{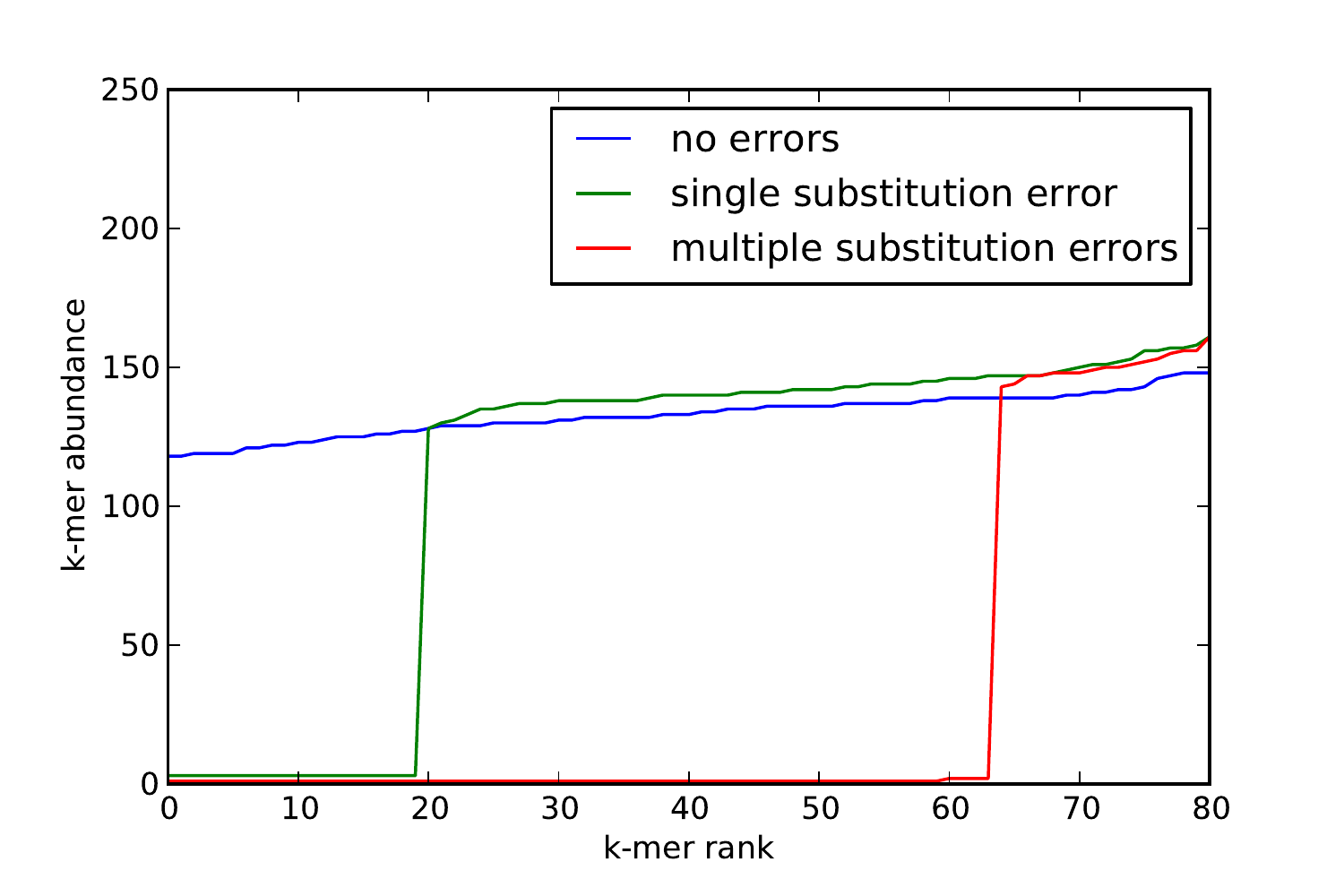}}
\caption{
{\bf Representative rank-abundance distributions for 20-mers from 100-base reads with no errors,
a read with a single substitution error, and a read with multiple
substitution errors.}}
\label{fig:rankabund}
\end{figure}

We may calculate a {\em reference-free} estimate of genome coverage by
looking at the k-mer abundance distribution within individual reads.
First, observe that k-mers, DNA words of a fixed length $k$, tend to
have similar abundances within a read: this is a well-known property
of k-mers that stems from each read originating from a single source
molecule of DNA.  The more times a region is sequenced, the higher the
abundance of k-mers from that region would be.  In the absence of
errors, average k-mer abundance could be used as an estimate of the
depth of coverage for a particular read (Figure \ref{fig:rankabund},
``no errors'' line).  However, when reads contain random substitution
or indel errors from sequencing, the k-mers overlapping these errors
will be of lower abundance; this feature is often used in k-mer based
error correction approaches \cite{pubmed21114842}.  For example, a
single substitution will introduce $k$ low-abundance k-mers within a
read.  (Figure \ref{fig:rankabund}, ``single substitution error''
line).  However, for small $k$ and reads of length $L$ where $L >
3k-1$, a single substitution error will not skew the {\em median}
k-mer abundance.  Only when multiple substitution errors are found in
a single read will the median k-mer abundance be affected (Figure
\ref{fig:rankabund}, ``multiple substitution errors'').

\begin{figure}[!ht]
\begin{center}
\centerline{\includegraphics[width=3in]{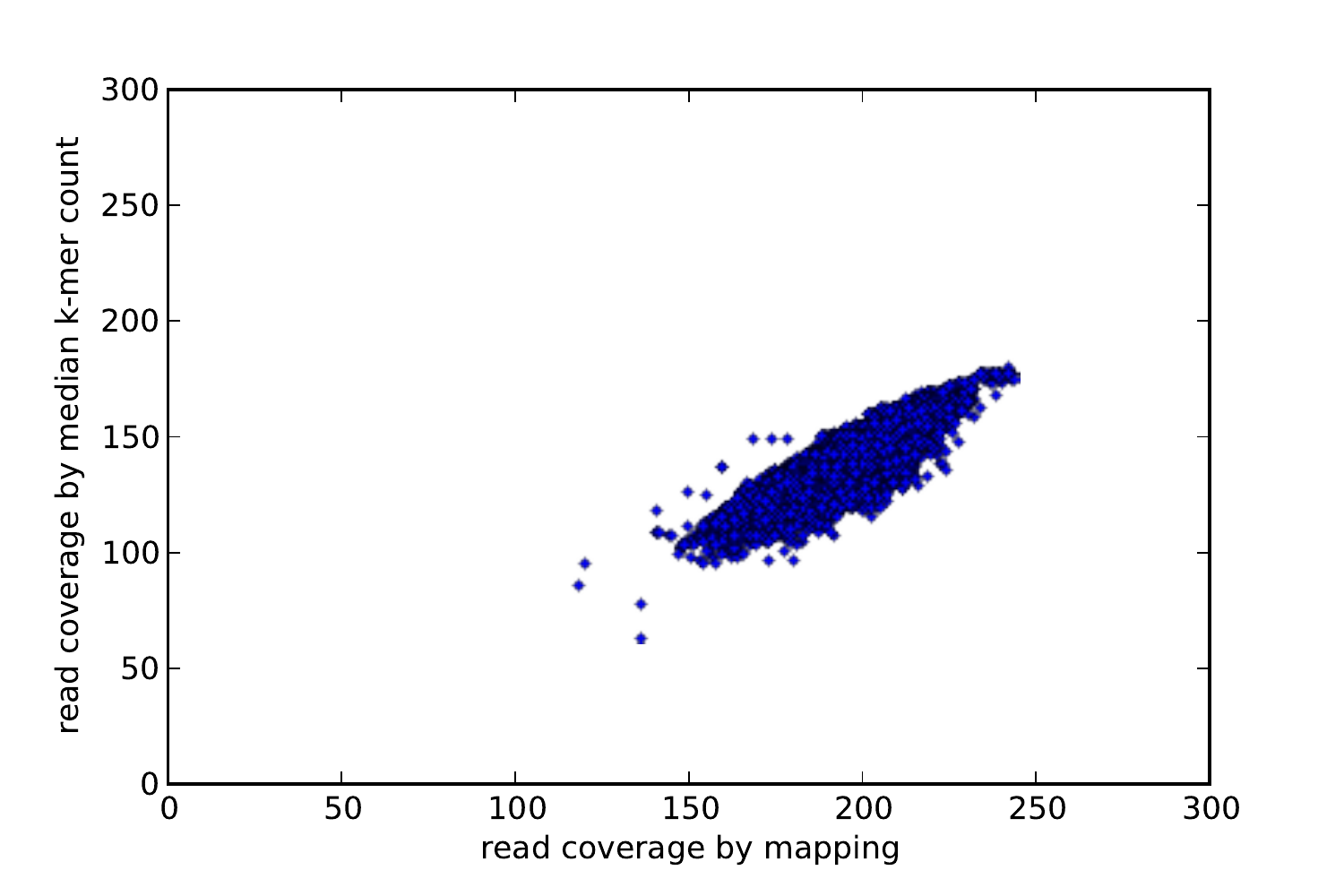}}
\centerline{\includegraphics[width=3in]{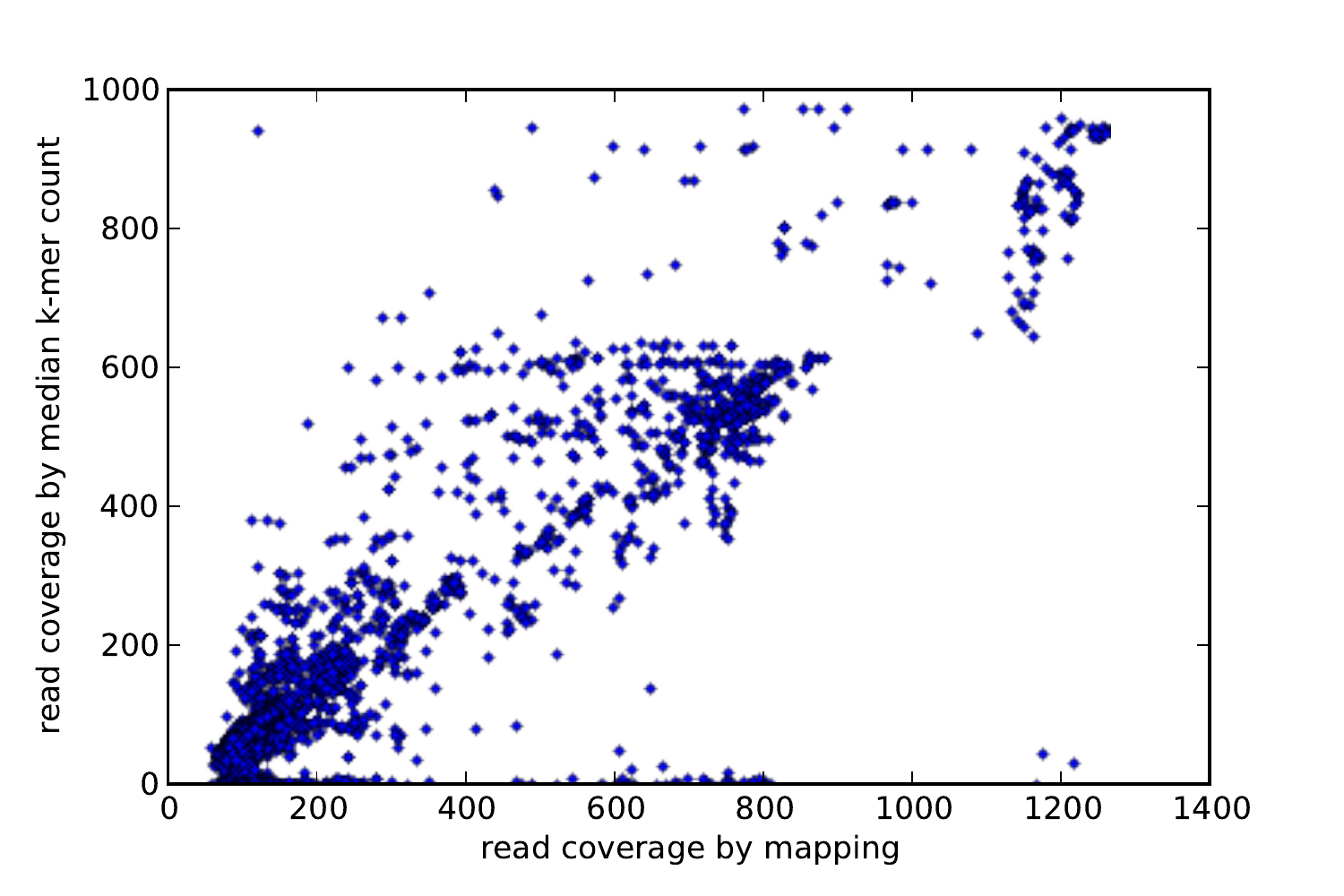}}
\end{center}
\caption{
{\bf Mapping and k-mer coverage measures correlate for simulated genome
data and a real {\em E. coli} data set (5m reads).  Simulated data $r^2 = 0.79$; {\em
E. coli} $r^2 = 0.80$.}
}
\label{fig:random}
\end{figure}

\begin{figure}[!ht]
\begin{center}
\centerline{\includegraphics[width=3in]{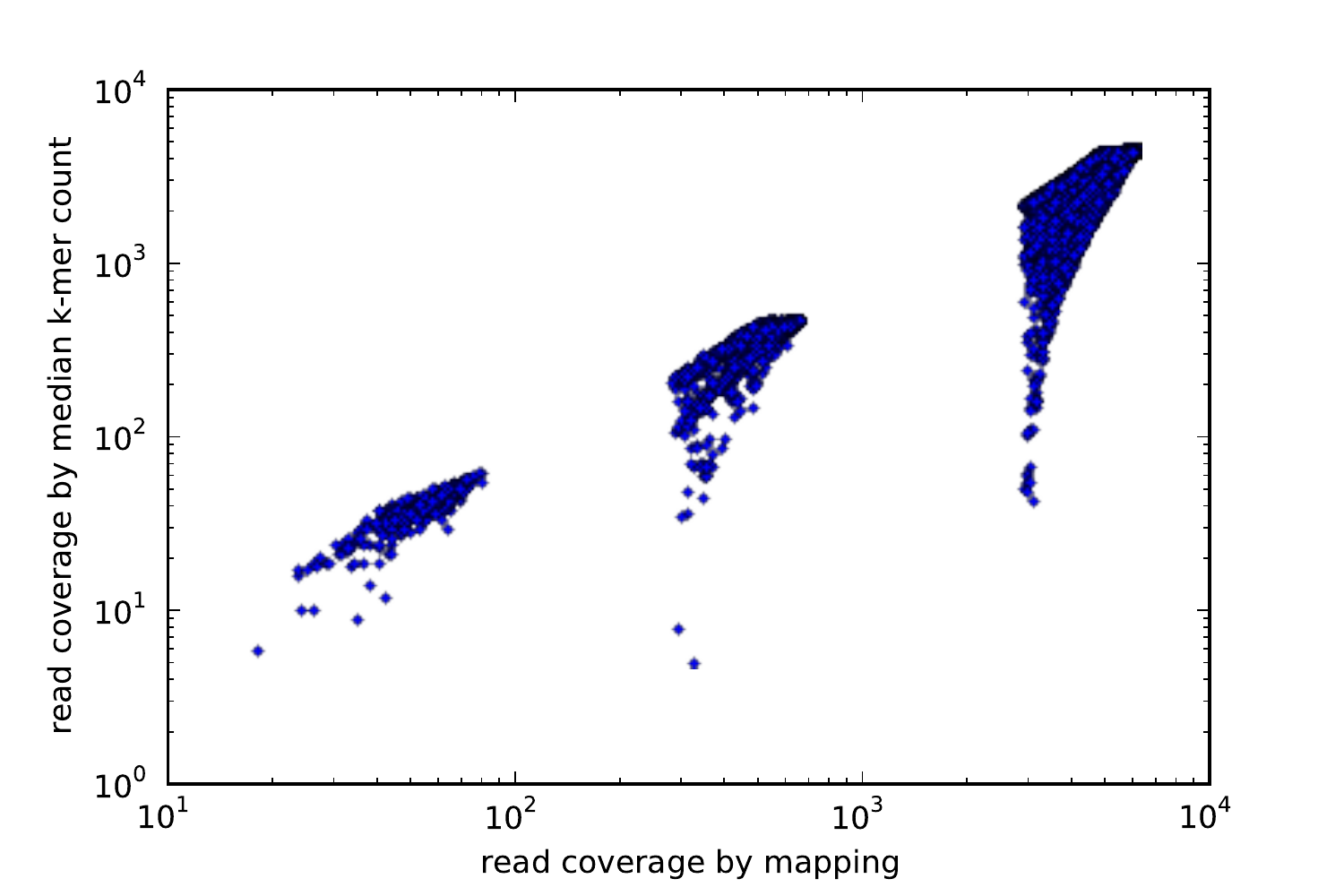}}
\centerline{\includegraphics[width=3in]{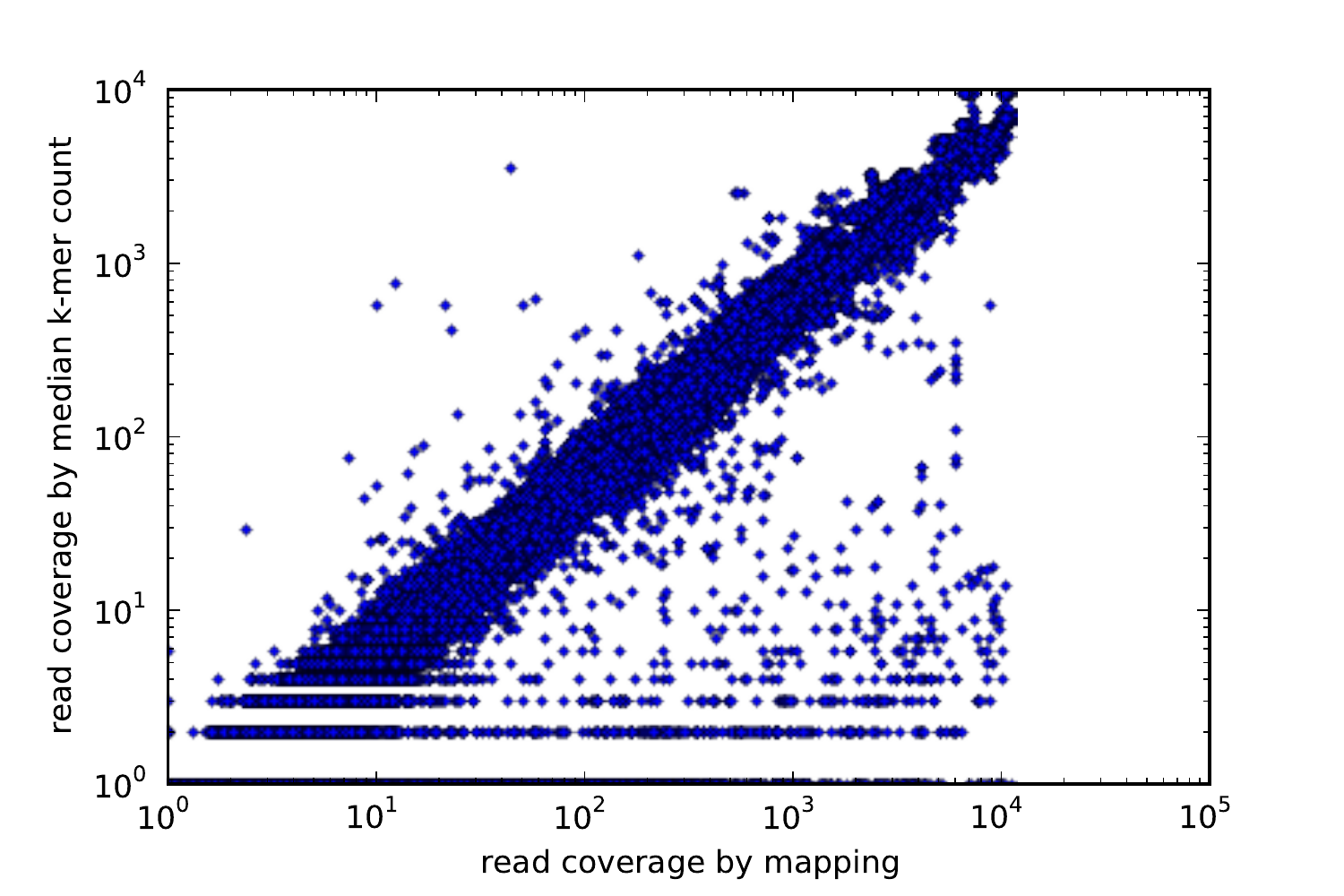}}
\end{center}
\caption{
{\bf Mapping and k-mer coverage measures correlate for simulated transcriptome data as well as real mouse transcriptome data. Simulated data $r^2 = 0.93$;
mouse transcriptome $r^2 = 0.90$.}
}
\label{fig:transcripts}
\end{figure}

Using a fixed-memory CountMin Sketch data structure to count k-mers
(see Methods and \cite{countminsketch}), we find that median k-mer
abundance correlates well with mapping-based coverage for artificial
and real genomic data sets.  There is a strong correlation between
median k-mer abundance and mapping-based coverage both for simulated
100-base reads generated with 1\% error from a 400kb artificial genome
sequence ($r^2 = 0.79$; also see Figure \ref{fig:random}a), as well as
for real short-read data from {\em E. coli} ($r^2 = 0.80$, also see
Figure \ref{fig:random}b).  This correlation also holds for simulated
and real mRNAseq data: for simulated transcriptome data, $r^2 = 0.93$
(Figure \ref{fig:transcripts}a), while for real mouse transcriptome
data, $r^2 = 0.90$ (Figure \ref{fig:transcripts}b).
Thus the median k-mer abundance of a read correlates
well with mapping-based estimates of read coverage.


\subsection*{Eliminating redundant reads reduces variation in sequencing depth}

Deeply sequenced genomes contain many highly covered loci.  For
example, in a human genome sequenced to 100x average coverage, we would
expect 50\% or more of the reads to have a coverage greater than 100.
In practice, we need many fewer of these reads to assemble
the source locus.

Using the median k-mer abundance estimator discussed above, we can examine each read
in the data set progressively to determine if it is high coverage.  At
the beginning of a shotgun data set, we would expect many reads to be
entirely novel and have a low estimated coverage.  As we proceed
through the data set, however, average coverage will increase and many
reads will be from loci that we have already sampled sufficiently.

Suppose we choose a coverage threshold $C$ past which we no longer
wish to collect reads. If we only keep reads whose estimated coverage
is less than $C$, and discard the rest, we will reduce the average
coverage of the data set to $C$.  This procedure is
algorithmically straightforward to execute: we examine each read's
estimated coverage, and retain only those whose coverage is less than $C$.
The following pseudocode provides one approach:
\begin{verbatim}
   for read in dataset:
      if estimated_coverage(read) < C:
         accept(read)
      else:
         discard(read)
\end{verbatim}
\noindent
where accepted reads contribute to the $\tt estimated\_coverage$
function.  Note that for any data set with an average coverage $> 2C$,
this has the effect of discarding the majority of reads.  Critically,
low-coverage reads, especially reads from undersampled regions, will
always be retained.

\begin{figure}[!ht]
\centerline{\includegraphics[width=4in]{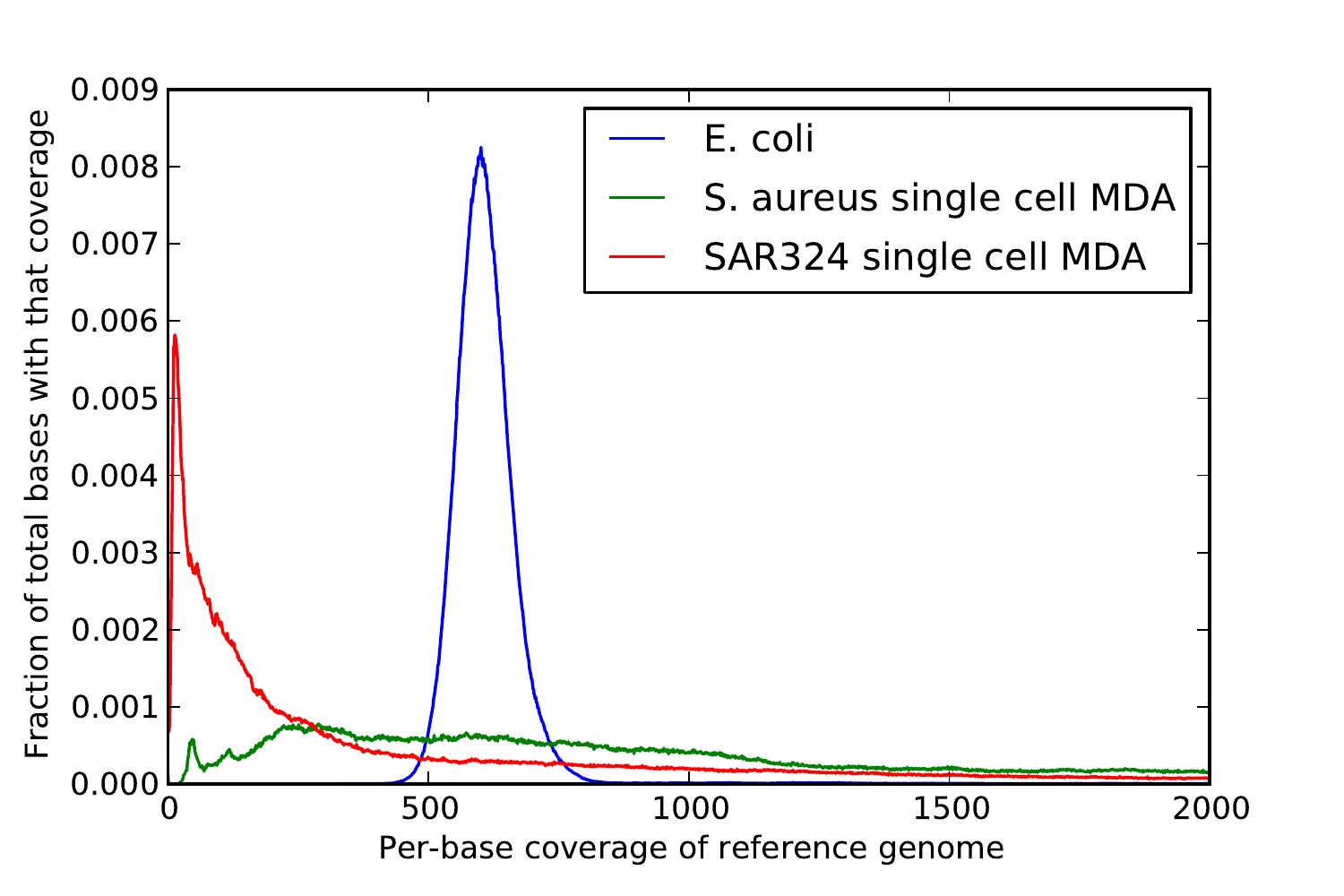}}
\centerline{\includegraphics[width=4in]{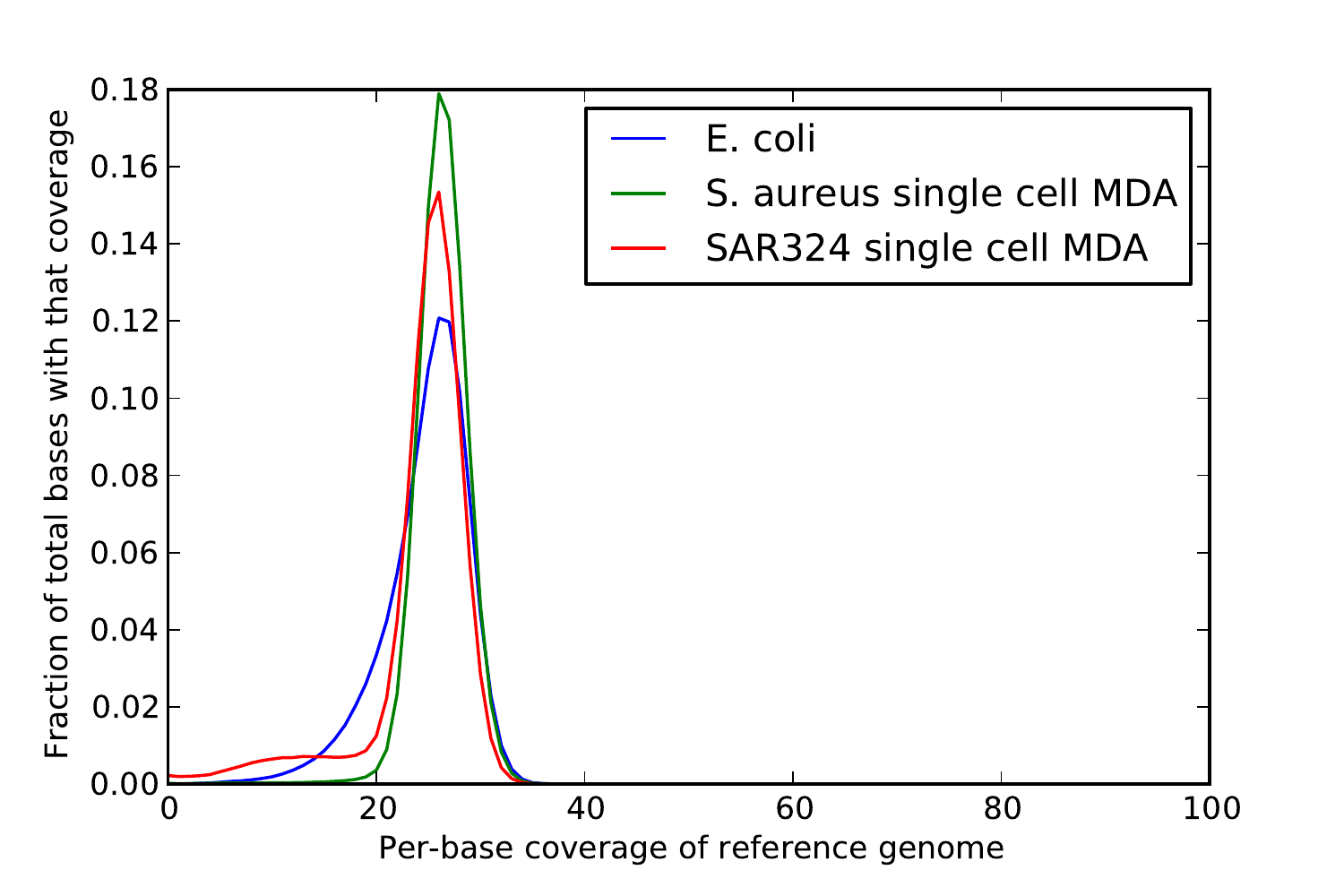}}
\caption{
{\bf Coverage distribution of three microbial genome samples, calculated
from mapped reads (a) before and (b) after digital normalization (k=20, C=20).}}
\label{fig:coverage}
\end{figure}

The net effect of this procedure, which we call digital normalization,
is to normalize the coverage distribution of data sets.  In Figure
\ref{fig:coverage}a, we display the estimated coverage of an {\em
  E. coli} genomic data set, a {\em S. aureus} single-cell
MD-amplified data set, and an MD-amplified data set from an uncultured
{\em Deltaproteobacteria}, calculated by mapping reads to the known or
assembled reference genomes (see \cite{pubmed21926975} for the data
source).  The wide variation in coverage for the two MDA data sets is
due to the amplification procedure \cite{pubmed17487184}.  After
normalizing to a k-mer coverage of 20, the high coverage loci are
systematically shifted to an average mapping coverage of 26, while
lower-coverage loci remain at their previous coverage.  This smooths
out coverage of the overall data set.

\begin{figure}[!ht]
\centerline{\includegraphics[width=4in]{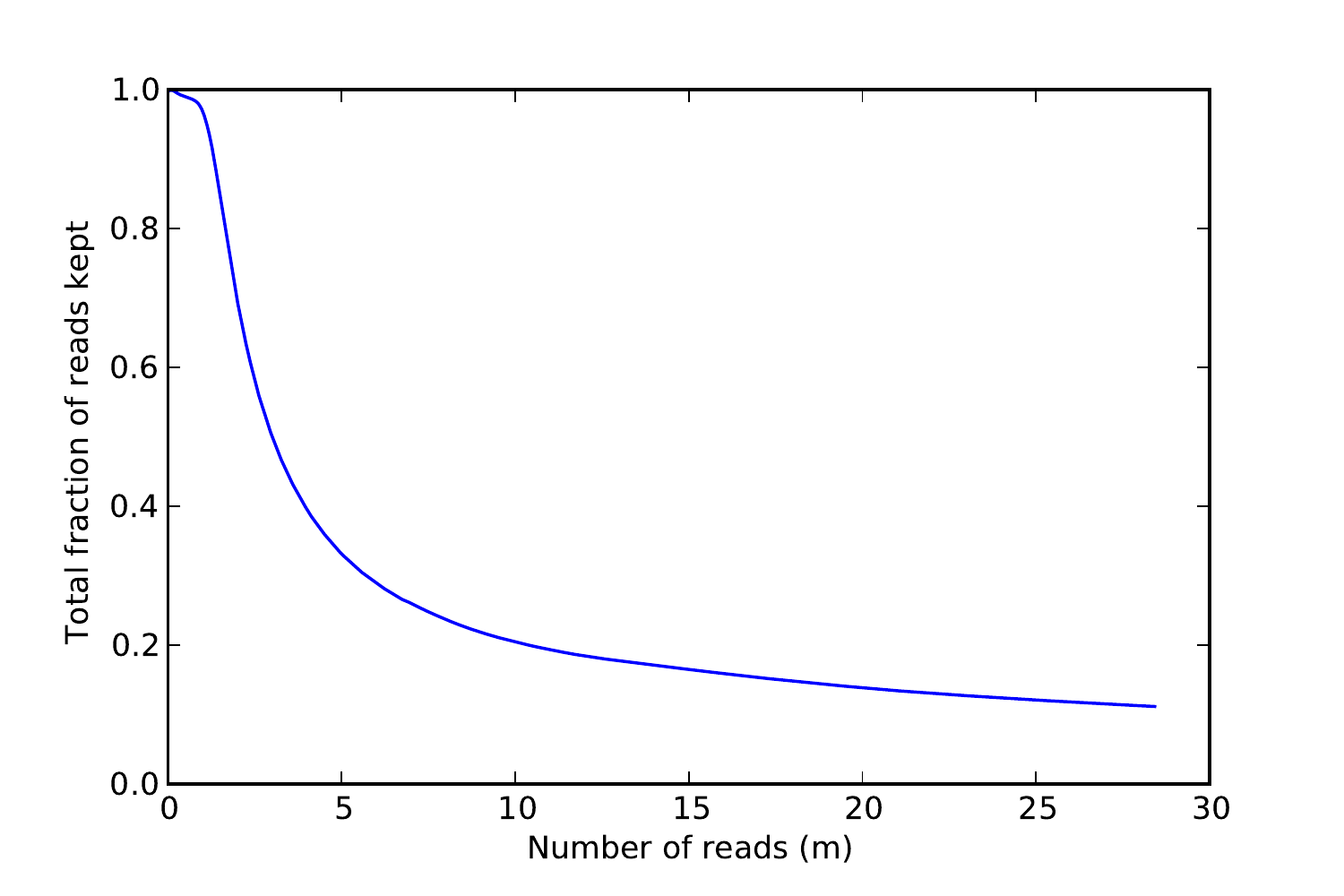}}
\caption{
{\bf Fraction of reads kept when normalizing the {\em E. coli} dataset to C=20 at k=20.}}
\label{fig:accumulate}
\end{figure}

At what rate are sequences retained?  For the {\em E. coli} data set,
Figure \ref{fig:accumulate} shows the fraction of sequences retained
by digital normalization as a function of the total number of reads
examined when normalizing to C=20 at k=20.  There is a clear
saturation effect showing that as more reads are examined, a smaller
fraction of reads is retained; by 5m reads, approximately 50-100x
coverage of {\em E. coli}, under 30\% of new reads are kept.  This
demonstrates that as expected, only a small amount of novelty (in
the form of either new information, or the systematic accumulation of
errors) is being observed with increasing sequencing depth.

\subsection*{Digital normalization retains information while discarding
both data and errors}

The 1-2\% per-base error rate of next-generation sequencers
dramatically affect the total number of k-mers.  For example, in the
simulated genomic data of 200x, a 1\% error rate leads to
approximately 20 new k-mers for each error, yielding 20-fold more
k-mers in the reads than are truly present in the genome (Table 1, row
1).  This in turn dramatically increases the memory requirements for
tracking and correcting k-mers \cite{pubmed21245053}.  This is a
well-known problem with de Bruijn graph approaches, in which erroneous
nodes or edges quickly come to dominate deep sequencing data sets.

When we perform digital normalization on such a data set, we eliminate
the vast majority of these k-mers (Table \ref{tab:normC20}, row 1).
This is because we are accepting or rejecting entire reads; in
going from 200x random coverage to 20x systematic coverage, we
discard 80\% of the reads containing 62\% of the errors (Table
\ref{tab:normC20}, row 1).  For reads taken from a skewed abundance
distribution, such as with MDA or mRNAseq, we similarly discard many
reads, and hence many errors (Table \ref{tab:normC20}, row 2).  In
fact, in most cases the process of sequencing fails to recover
more true k-mers (Table \ref{tab:normC20}, middle column, parentheses) than
digital
normalization discards (Table \ref{tab:normC20}, fourth column, parentheses).

The net effect of digital normalization is to retain nearly all {\em
  real} k-mers, while discarding the majority of erroneous k-mers --
in other words, digital normalization is discarding {\em data} but not
{\em information}.  This rather dramatic elimination of erroneous
k-mers is a consequence of the high error rate present in reads: with
a 1\% per-base substitution error rate, each 100-bp read will have an
average of one substitution error. Each of these substitution errors
will introduce up to $k$ erroneous k-mers.  Thus, for each read we
discard as redundant, we also eliminate an average of $k$ erroneous
k-mers.

We may further eliminate erroneous k-mers by removing k-mers that are
rare across the data set; these rare k-mers tend to result from
substitution or indel errors \cite{pubmed21114842}.  We do this by
first counting all the k-mers in the accepted reads during digital
normalization.  We then execute a second pass across the accepted
reads in which we eliminate the 3' ends of reads at low-abundance
k-mers.  Following this error reduction pass, we execute a
second round of digital normalization (a third pass across the data
set) that further eliminates redundant data.  This three-pass protocol
eliminates additional errors and results in a further decrease in data
set size, at the cost of very few real k-mers in genomic data sets
(Table \ref{tab:normC5}).

Why use this three-pass protocol rather than simply normalizing to the
lowest desired coverage in the first pass?  We find that removing
low-abundance k-mers after a single normalization pass to $C \approx
5$ removes many more {\em real} k-mers, because there will be many
regions in the genome that by chance have yielded 5 reads with errors
in them. If these erroneous k-mers are removed in the abundance-trimming step,
coverage of the corresponding regions is eliminated.  By normalizing
to a higher coverage of 20, removing errors, and only then reducing
coverage to 5, digital normalization can retain accurate reads for most
regions.  Note that this three-pass protocol is not considerably more
computationally expensive than the single-pass protocol: the first
pass discards the majority of data and errors, so later passes are
less time and memory intensive than the first pass.

\begin{figure}[!ht]
\centerline{\includegraphics[width=4in]{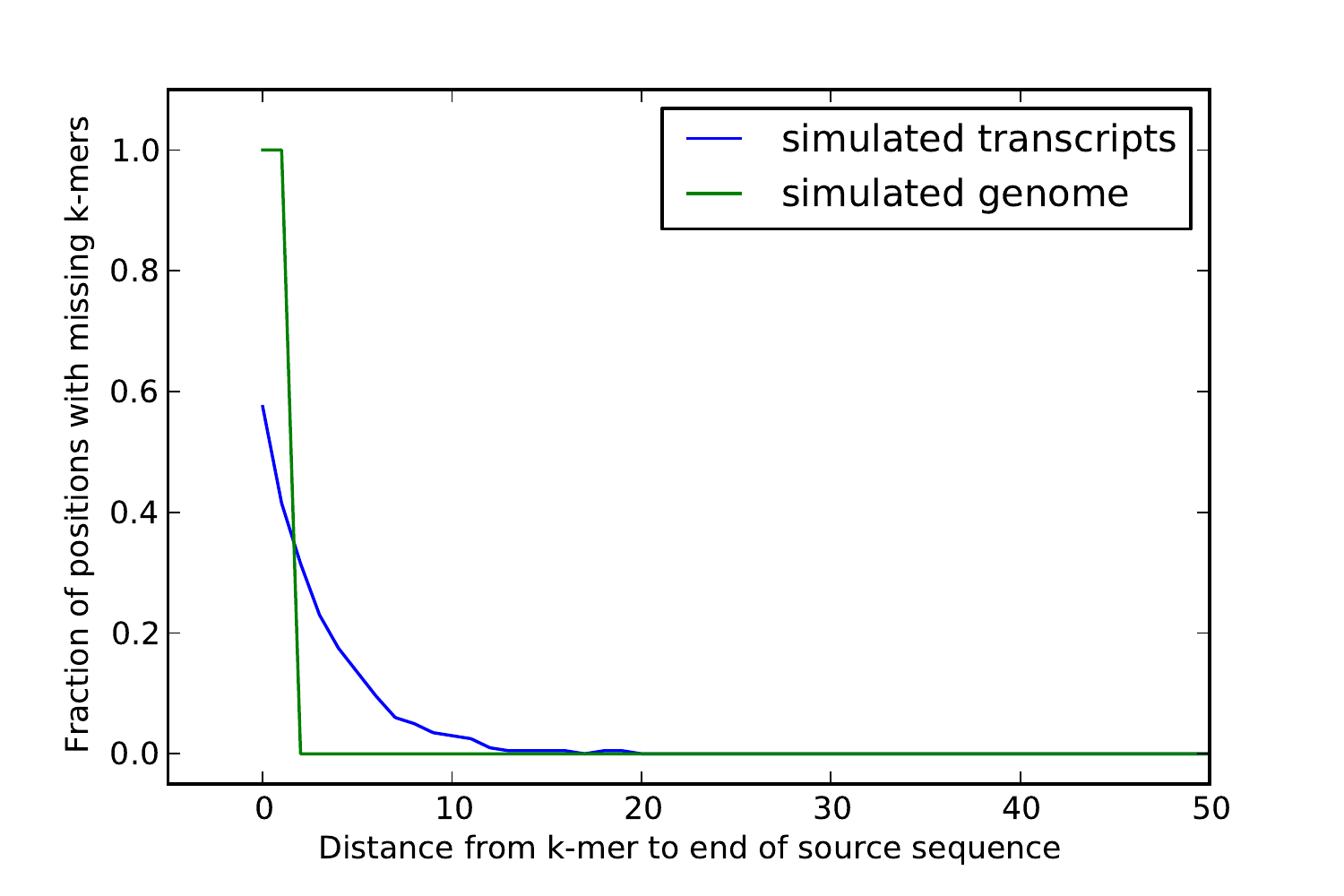}}
\caption{
{\bf K-mers at the ends of sequences are lost during digital normalization.}}
\label{fig:endloss}
\end{figure}

Interestingly, this three-pass protocol removed many more real k-mers
from the simulated mRNAseq data than from the simulated genome -- 351
of 48,100 (0.7\%) real k-mers are lost from the mRNAseq, vs 4 of
399,981 lost (.000001\%) from the genome (Table \ref{tab:normC5}).
While still only a tiny fraction of the total number of real k-mers,
the difference is striking -- the simulated mRNAseq sample loses k-mers
at almost 1000-fold the rate of the simulated genomic sample.  Upon
further investigation, all but one of the lost k-mers were located
within 20 bases of the ends of the source sequences; see Figure
\ref{fig:endloss}.  This is because digital normalization cannot
distinguish between erroneous k-mers and k-mers that are undersampled
due to edge effects.  In the case of the simulated genome, which was
generated as one large chromosome, the effect is negligible, but the
simulated transcriptome was generated as 100 transcripts of length
500.  This added 99 end sequences over the genomic simulation, which
in turn led to many more lost k-mers.

While the three-pass protocol is effective at removing erroneous
k-mers, for some samples it may be too stringent.  For example, the
mouse mRNAseq data set contains only 100m reads, which may not be
enough to thoroughly sample the rarest molecules; in this case the
abundance trimming would remove real k-mers as well as erroneous k-mers.
Therefore we used the single-pass digital normalization
for the yeast and mouse transcriptomes.  For these two samples we can
also see that the first-pass digital normalization is extremely effective,
eliminating essentially all of the erroneous k-mers (Table \ref{tab:normC20},
rows 4 and 5.)

\subsection*{Digital normalization scales assembly of microbial genomes}

We applied the three-pass digital normalization and error trimming
protocol to three real data sets from Chitsaz et al (2011)
\cite{pubmed21926975}.  The first pass of digital normalization was
performed in 1gb of memory and took about 1 min per million reads.
For all three samples, the number of reads remaining after digital
normalization was reduced by at least 30-fold, while the memory and
time requirements were reduced 10-100x.


Despite this dramatic reduction in data set size and computational
requirements for assembly, both the {\em E. coli} and {\em S. aureus}
assemblies overlapped with the known reference sequence by more than
98\%.  This confirms that little or no information was lost during
the process of digital normalization; moreover, it appears that
digital normalization does not significantly affect the assembly results.
(Note that we did not perform scaffolding, since the digital
normalization algorithm does not take into account paired-end
sequences, and could mislead scaffolding approaches.  Therefore, these
results cannot directly be compared to those in Chitsaz et al. (2011)
\cite{pubmed21926975}.)

The {\em Deltaproteobacteria} sequence also assembled well, with
98.8\% sequence overlap with the results from Chitsaz et al.
Interestingly, only 30kb of the sequence assembled with Velvet-SC in
Chitsaz et al. (2011) was missing, while an additional 360kb of
sequence was assembled only in the normalized samples.  Of the 30kb of
missing sequence, only 10\% matched via TBLASTX to a nearby {\em
  Deltaproteobacteria} assembly, while more than 40\% of the
additional 360kb matched to the same {\em Deltaproteobacteria} sample.
Therefore these additional contigs likely represents real
sequence, suggesting that digital normalization is competitive with
Velvet-SC in terms of sensitivity.



\subsection*{Digital normalization scales assembly of transcriptomes}

We next applied single-pass digital normalization to published yeast
and mouse mRNAseq data sets, reducing them to 20x coverage at k=20
\cite{pubmed21572440}.  Digital normalization on these samples used
8gb of memory and took about 1 min per million reads.  We then
assembled both the original and normalized sequence reads with Oases
and Trinity, two {\em de novo} transcriptome assemblers (Table
\ref{tab:dntrans}) \cite{pubmed22368243,pubmed21572440}.

For both assemblers the computational resources necessary to complete
an assembly were reduced (Table \ref{tab:dntrans}), but normalization
had different effects on performance for the different samples.  On the
yeast data set, time and memory requirements were reduced
significantly, as for Oases running on mouse.  However, while
Trinity's runtime decreased by a factor of three on the normalized
mouse data set, the memory requirements did not decrease
significantly.  This may be because the mouse transcriptome is 5-6
times larger than the yeast transcriptome, and so the mouse mRNAseq
is lower coverage overall; in this case we would expect fewer
errors to be removed by digital normalization.

The resulting assemblies differed in summary statistics (Table
\ref{tab:dntrans0}).  For both yeast and mouse, Oases lost 5-10\% of
total transcripts and total bases when assembling the normalized data.  However, Trinity {\em gained}
transcripts when assembling the normalized yeast and mouse data,
gaining about 1\% of total bases on yeast and losing about 1\%
of total bases in mouse.  Using a local-alignment-based overlap
analysis (see Methods) we found little difference in sequence
content between the pre- and post- normalization assemblies: for
example, the normalized Oases assembly had a 98.5\% overlap with the
unnormalized Oases assembly, while the normalized Trinity assembly had
a 97\% overlap with the unnormalized Trinity assembly.

To further investigate the differences between transcriptome
assemblies caused by digital normalization, we looked at the
sensitivity with which long transcripts were recovered
post-normalization.  When comparing the normalized assembly to the
unnormalized assembly in yeast, Trinity lost only 3\% of the sequence
content in transcripts greater than 300 bases, but 10\% of the
sequence content in transcripts greater than 1000 bases.  However,
Oases lost less than 0.7\% of sequence content at 300 and
1000 bases.  In mouse, we see the same pattern.
This suggests that the change in summary statistics for
Trinity is caused by fragmentation of long transcripts into shorter
transcripts, while the difference for Oases is caused by loss of
splice variants.  Indeed, this
loss of splice variants should be expected, as there are many low-prevalence splice
variants present in deep sequencing data \cite{pubmed21151575}.
Interestingly, in yeast we recover {\em more} transcripts after
digital normalization; these transcripts appear to be additional splice
variants.


The difference between Oases and Trinity results show that Trinity is
more sensitive to digital normalization than Oases: digital
normalization seems to cause Trinity to fragment long transcripts.
Why?  One potential issue is that Trinity only permits k=26 for
assembly, while normalization was performed at k=20; digital
normalization may be removing 26-mers that are important for Trinity's
path finding algorithm.  Alternatively, Trinity may be more sensitive
than Oases to the change in coverage caused by digital normalization.
Regardless, the strong performance of Oases on digitally normalized
samples, as well as the high retention of k-mers (Table \ref{tab:normC20})
suggests that the primary sequence content for the transcriptome remains
present in the normalized reads, although it is recovered with different
effectiveness by the two assemblers.

\section*{Discussion}

\subsection*{Digital normalization dramatically scales {\em de novo} assembly}

The results from applying digital normalization to read data sets
prior to {\em de novo} assembly are extremely good: digital
normalization reduces the computational requirements (time and memory)
for assembly considerably, without substantially affecting the
assembly results.  It does this in two ways: first, by removing
the majority of reads without significantly affecting the true k-mer
content of the data set. Second, by eliminating these reads,
digital normalization also eliminates sequencing errors contained
within those reads, which otherwise would add significantly to memory
usage in assembly \cite{pubmed21245053}.

Digital normalization also lowers computational requirements by
eliminating most repetitive sequence in the data set.
Compression-based approaches to graph storage have demonstrated that
compressing repetitive sequence also effectively reduces memory and
compute requirements \cite{pubmed22139935,pubmed22156294}.  Note
however that {\em eliminating} many repeats may also have its
negatives (discussed below).

Digital normalization should be an effective preprocessing approach
for most assemblers.  In particular, the de Bruijn graph approach used
in many modern assemblers relies on k-mer content, which is almost
entirely preserved by digital normalization (see Tables \ref{tab:normC20}
and \ref{tab:normC5}) \cite{pubmed20211242}.

\subsection*{A general strategy for normalizing coverage}

Digital normalization is a general strategy for systematizing coverage
in shotgun sequencing data sets by using per-locus downsampling,
albeit without any prior knowledge of reference loci.  This yields
considerable theoretical and practical benefits in the area of {\em de
  novo} sequencing and assembly.

In theoretical terms, digital normalization offers a general strategy
for changing the scaling behavior of sequence assembly.  Assemblers
tend to scale poorly with the number of reads: in particular, de
Bruijn graph memory requirements scale linearly with the size of the
data set due to the accumulation of errors, although others have
similarly poor scaling behavior (e.g. quadratic time in the number of
reads) \cite{pubmed20211242}.  By calculating per-locus coverage in a way that
is insensitive to errors, digital normalization converts
genome assembly into a problem that scales with the complexity of the
underlying sample - i.e. the size of the genome, transcriptome, or
metagenome.

Digital normalization also provides a general strategy for applying
online or streaming approaches to analysis of {\em de novo} sequencing
data.  The basic algorithm presented here is explicitly a single-pass or streaming
algorithm, in which the entire data set is never considered as a
whole; rather, a partial ``sketch'' of the data set is retained and
used for progressive filtering.  Online algorithms and sketch data
structures offer significant opportunities in situations where data
sets are too large to be conveniently stored, transmitted, or analyzed
\cite{muthukrishnan2005data}.  This can enable increasingly efficient
downstream analyses.
Digital normalization can be applied in any situation where the
abundance of particular sequence elements is either unimportant or can be
recovered more efficiently after other processing, as in assembly.

The construction of a simple, reference-free measure of coverage on a
per-read basis offers opportunities to analyze coverage and
diversity with an assembly-free approach.  Genome and transcriptome
sequencing is increasingly being applied to non-model organisms and
ecological communities for which there are no reference sequences, and
hence no good way to estimate underlying sequence complexity.  The
reference-free counting technique presented here provides a method for
determining community and transcriptome complexity;
it can also be used to progressively estimate sequencing depth.

More pragmatically, digital normalization also scales existing
assembly techniques dramatically.
The reduction in data set size afforded by
digital normalization may also enable the application of more
computationally expensive algorithms such as overlap-layout-consensus
assembly approaches to short-read data.  Overall, the reduction in
data set size, memory requirements, and time complexity for contig
assembly afforded by digital normalization could lead to the
application of more complex heuristics to the assembly problem.

\subsection*{Digital normalization drops terminal k-mers and removes isoforms}

Our implementation of digital normalization does discard some real
information, including terminal k-mers and low-abundance isoforms.
Moreover, we predict a number of other failure modes: for example,
because k-mer approaches demand strict sequence identity, data sets
from highly polymorphic organisms or populations will perform more
poorly than data sets from low-variability samples.  Digital
normalization also discriminates against highly repetitive
sequences. We note that these problems traditionally have been
challenges for assembly strategies: recovering low-abundance isoforms
from mRNAseq, assembling genomes from highly polymorphic organisms,
and assembling across repeats are all difficult tasks, and
improvements in these areas continue to be active areas of research
\cite{pubmed18549302,pubmed20633259,pubmed18541131}.  Using an
alignment-based approach to estimating coverage, rather than a k-mer
based approach, could provide an alternative implementation that would
improve performance on errors, splice variants, and terminal k-mers.
Our current approach also ignores quality scores; a ``q-mer'' counting
approach as in Quake, in which k-mer counts are weighted by quality
scores, could easily be adapted \cite{pubmed21114842}.

Another concern for normalizing deep sequencing data sets is that,
with sufficiently deep sequencing, sequences with many errors will
start to accrue.  This underlies the continued accumulation of
sequence data for {\em E. coli} observed in Figure
\ref{fig:accumulate}.  Assemblers may be unable to distinguish between
this false sequence and the error-free sequences, for sufficiently
deep data sets.  This accumulation of erroneous sequences is again
caused by the use of k-mers to detect similarity, and is one reason
why exploring local alignment approaches (discussed below) may be a
good future direction.

\newpage

\subsection*{Applying assembly algorithms to digitally normalized data}


The assembly problem is challenging for several reasons: many
formulations are computationally complex (NP-hard), and practical
issues of both genome content and sequencing, such as repetitive
sequence, polymorphisms, short reads and high error rates, challenge
assembly approaches \cite{pubmed19580519}.  This has driven the
development of heuristic approaches to resolving complex regions in
assemblies.  Several of these heuristic approaches use the abundance
information present in the reads to detect and resolve repeat regions;
others use pairing information from paired-end and mate-pair sequences
to resolve complex paths.  Digital normalization aggressively removes
abundance information, and we have not yet adapted it to paired-end
sequencing data sets; this could and should affect the quality of
assembly results! Moreover, it is not clear what effect different
coverage (C) and k-mer (k) values have on assemblers.  In practice,
for at least one set of k-mer size $k$ and normalized coverage $C$
parameters, digital normalization seems to have little negative effect
on the final assembled contigs.  Further investigation of the effects
of varying $k$ and $C$ relative to specific assemblers and assembler
parameters will likely result in further improvements in assembly
quality.

%

A more intriguing notion than merely using digital normalization as a
pre-filter is to specifically adapt assembly algorithms and protocols
to digitally normalized data.  For example, the reduction in data set
size afforded by digital normalization may make
overlap-layout-consensus approaches computationally feasible for
short-read data \cite{pubmed20211242}.  Alternatively, the quick and
inexpensive generation of contigs from digitally normalized data could
be used prior to a separate scaffolding step, such as those supported
by SGA and Bambus2 \cite{pubmed20529929,pubmed21926123}.  Digital
normalization offers many future directions for improving assembly.

\subsection*{Conclusions}

Digital normalization is an effective demonstration
that much of short-read shotgun sequencing is redundant.  Here we have
shown this by normalizing samples to 5-20x coverage while recovering
complete or nearly complete contig assemblies.  Normalization is
substantially different from uniform downsampling: by doing
downsampling in a locus-specific manner, we retain low coverage data.
Previously described approaches to reducing sampling variation rely on
{\em ad hoc} parameter measures and/or an initial round of assembly
and have not been shown to be widely applicable
\cite{pubmed19724646,pubmed22028825}.



We have implemented digital normalization as a {\em prefilter} for assembly, so
that any assembler may be used on the normalized data.  Here we have
only benchmarked a limited set of assemblers -- Velvet, Oases, and
Trinity -- but in theory digital normalization should apply to any
assembler.  De Bruijn and string graph assemblers such as Velvet, SGA,
SOAPdenovo, Oases, and Trinity are especially likely to work well with
digital normalized data, due to the underlying reliance on k-mer
overlaps in these assemblers.



\subsection*{Digital normalization is widely applicable and computationally convenient}
Digital normalization can be applied {\em de novo} to {\em any}
shotgun data set.  The approach is extremely computationally
convenient: the runtime complexity is linear with respect to the data
size, and perhaps more importantly it is {\em single-pass}: the basic
algorithm does not need to look at any read more than once.  Moreover,
because reads accumulate sub-linearly, errors do not accumulate
quickly and overall memory requirements for digital normalization
should grow slowly with data set size.  Note also that while the
algorithm presented here is not perfectly parallelizable, efficient
distributed k-mer counting is straightforward and it should be
possible to scale digital normalization across multiple machines
\cite{pubmed19357099}.

The first pass of digital normalization is implemented as an online
streaming algorithm in which reads are examined once.  Streaming
algorithms are useful for solving data analysis problems in which
the data are too large to easily be transmitted, processed, or
stored.  Here, we implement the streaming algorithm using a fixed
memory data ``sketch'' data structure, CountMin Sketch.  By combining
a single-pass algorithm with a fixed-memory data structure, we provide
a data reduction approach for sequence data analysis with both
(linear) time and (constant) memory guarantees. Moreover, because the
false positive rate of the CountMin Sketch data structure is well
understood and easy to predict, we can provide {\em data quality}
guarantees as well.  These kinds of guarantees are immensely
valuable from an algorithmic perspective, because they provide a
robust foundation for further work \cite{muthukrishnan2005data}.
%
The general concept of removing redundant {\em data} while retaining
{\em information} underlies ``lossy compression'', an approach used
widely in image processing and video compression.  The concept of
lossy compression has broad applicability in sequence analysis.
For example, digital normalization could be applied prior to homology
search on unassembled reads, potentially reducing the computational
requirements for e.g. BLAST and HMMER without loss of sensitivity.
Digital normalization could also help merge multiple different read
data sets from different read technologies, by discarding
entirely redundant sequences and retaining only sequences containing
``new'' information.  These approaches remain to be explored in the future.






\section*{Methods}

All links below are available electronically through
ged.msu.edu/papers/2012-diginorm/, in addition to the
archival locations provided.

\subsection*{Data sets}

The {\em E. coli}, {\em S. aureus}, and {\em Deltaproteobacteria} data
sets were taken from Chitsaz et al. \cite{pubmed21926975}, and
downloaded from bix.ucsd.edu/projects/singlecell/.  The
mouse data set was published by Grabherr et al. \cite{pubmed21572440}
and downloaded from trinityrnaseq.sf.net/.  All data sets
were used without modification.
The complete assemblies, both pre- and post-normalization, for the
{\em E. coli}, {\em S. aureus}, the uncultured {\em
  Deltaproteobacteria}, mouse, and yeast data sets are available from
ged.msu.edu/papers/2012-diginorm/.

The simulated genome and transcriptome were generated from a uniform
AT/CG distribution.  The genome consisted of a single chromosome
400,000 bases in length, while the transcriptome consisted of 100
transcripts of length 500.  100-base reads were generated uniformly
from the genome to an estimated coverage of 200x, with a random 1\%
per-base error.  For the transcriptome, 1 million reads of length 100
were generated from the transcriptome at relative expression levels of
10, 100, and 1000, with transcripts assigned randomly with equal
probability to each expression group; these reads also had a 1\%
per-base error.

\subsection*{Scripts and software}

All simulated data sets and all analysis summaries were generated by
Python scripts, which are available at
github.com/ged-lab/2012-paper-diginorm/.  Digital normalization and k-mer
analyses were performed with the khmer software package, written in
C++ and Python, available at github.com/ged-lab/khmer/, tag
'2012-paper-diginorm'.  khmer also relies on the screed package for
loading sequences, available at github.com/ged-lab/screed/, tag
'2012-paper-diginorm'.  khmer and screed are Copyright (c) 2010
Michigan State University, and are free software available for
distribution, modification, and redistribution under the BSD license.

Mapping was performed with bowtie v0.12.7 \cite{pubmed19261174}.
Genome assembly was done with velvet 1.2.01 \cite{pubmed18349386}.
Transcriptome assembly was done with velvet 1.1.05/oases 0.1.22 and
Trinity, head of branch on 2011.10.29.
Graphs and correlation coefficients were generated using matplotlib
v1.1.0, numpy v1.7, and ipython notebook v0.12 \cite{ipython}.  The
ipython notebook file and data analysis scripts necessary to generate
the figures are available at 
  github.com/ged-lab/2012-paper-diginorm/.


\subsection*{Analysis parameters}

The khmer software uses a CountMin Sketch data structure to count
k-mers, which requires a fixed memory allocation
\cite{countminsketch}.  In all cases the memory usage was fixed such
that the calculated false positive rate was below 0.01.  By default k
was set to 20.

Genome and transcriptome coverage was calculated by mapping all reads
to the reference with bowtie ({\tt -a --best --strata}) and then
computing the per-base coverage in the reference.  Read coverage was
computed by then averaging the per-base reference coverage for each
position in the mapped read; where reads were mapped to multiple
locations, a reference location was chosen randomly for computing
coverage.  Median k-mer counts were computed with khmer as described
in the text.  Artificially high counts resulting from long stretches
of Ns were removed after the analysis.
Correlations between median k-mer counts and mapping coverage were
computed using numpy.corrcoef; see calc-r2.py script.

\subsection*{Normalization and assembly parameters}

For Table \ref{tab:dngenome}, the assembly k parameter for Velvet was
k=45 for {\em E. coli}; k=41 for {\em S. aureus} single cell; and k=39
for {\em Deltaproteobacteria} single cell.  Digital normalization
on the three bacterial samples was done with {\tt -N 4 -x 2.5e8 -k 20},
consuming 1gb of memory.  Post-normalization k parameters for Velvet
assemblies were k=37 for {\em E. coli}, k=27 for {\em S. aureus}, and k=27 for {\em Deltaproteobacteria}.
For Table \ref{tab:dntrans}, the assembly k parameter for Oases was k=21 for yeast
and k=23 for mouse.  Digital normalization on both mRNAseq samples was done
with {\tt -N 4 -x 2e9 -k 20}, consuming 8gb of memory.  Assembly of the
single-pass normalized mRNAseq was done with Oases at k=21 (yeast) and k=19
(mouse).

\subsection*{Assembly overlap and analysis}

Assembly overlap was computed by first using NCBI BLASTN to build local
alignments for two assemblies, then filtering for matches with bit scores
$>$ 200, and finally computing the fraction of bases in each assembly
with at least one alignment.  Total fractions were normalized to
self-by-self BLASTN overlap identity to account for BLAST-specific
sequence filtering.
TBLASTX comparison of the {\em Deltaproteobacteria} SAR324 sequence
was done against another assembled SAR324 sequence, acc AFIA01000002.1.

\subsection*{Compute requirement estimation}

Execution time was measured using real time from Linux bash 'time'.
Peak memory usage was estimated either by the 'memusg' script from
stackoverflow.com, peak-memory-usage-of-a-linux-unix-process, included
in the khmer repository; or by the Torque queuing system monitor, for
jobs run on MSU's HPC system.  While several different machines were
used for analyses, comparisons between unnormalized and normalized
data sets were always done on the same machine.

\section*{Acknowledgments}

We thank Chris Hart, James M. Tiedje, Brian Haas, Jared Simpson, Scott
Emrich, and Russell Neches for their insight and helpful technical
commentary.

\bibliography{diginorm}

\section*{Figure Legends}

\section*{Tables}

\begin{table}[!ht]
\caption{
\bf{Digital normalization to C=20 removes many erroneous k-mers from sequencing data sets.  Numbers
in parentheses indicate number of true k-mers lost at each step, based on reference.}}
\begin{tabular}{|l|c|c|c|c|c|}
Data set & True 20-mers & 20-mers in reads & 20-mers at C=20 & \% reads kept\\
\hline \\
Simulated genome & 399,981 & 8,162,813 & 3,052,007 (-2) & 19\% \\
Simulated mRNAseq & 48,100 & 2,466,638 (-88) & 1,087,916 (-9) & 4.1\% \\
{\em E. coli} genome & 4,542,150 & 175,627,381 (-152) & 90,844,428 (-5) & 11\% \\
Yeast mRNAseq & 10,631,882 & 224,847,659 (-683) & 10,625,416 (-6,469) & 9.3\% \\
Mouse mRNAseq & 43,830,642 & 709,662,624 (-23,196) & 43,820,319 (-13,400) & 26.4\% \\
\end{tabular}
\begin{flushleft}
\end{flushleft}
\label{tab:normC20}
\end{table}



\begin{table}[!ht]
\caption{
\bf{Three-pass digital normalization removes most erroneous k-mers.  Numbers
in parentheses indicate number of true k-mers lost at each step, based on known reference.}}
\begin{tabular}{|l|c|c|c|c|}
Data set & True 20-mers & 20-mers in reads & 20-mers remaining & \% reads kept\\
\hline \\
Simulated genome & 399,981 & 8,162,813 & 453,588 (-4) & 5\% \\
Simulated mRNAseq & 48,100 & 2,466,638 (-88) & 182,855 (-351) & 1.2\% \\
{\em E. coli} genome & 4,542,150 & 175,627,381 (-152) & 7,638,175 (-23) & 2.1\% \\
Yeast mRNAseq & 10,631,882 & 224,847,659 (-683) & 10,532,451 (-99,436) & 2.1\% \\
Mouse mRNAseq & 43,830,642 & 709,662,624 (-23,196) & 42,350,127 (-1,488,380) & 7.1\% \\
\end{tabular}
\begin{flushleft}
\end{flushleft}
\label{tab:normC5}
\end{table}


\begin{table}[!ht]
\caption{
\bf{Three-pass digital normalization reduces computational requirements for contig assembly of genomic data.}}
\begin{tabular}{|l|c|c|c|c|}

Data set & N reads pre/post & Assembly time pre/post & Assembly memory pre/post \\
\hline \\
{\em E. coli} & 31m / 0.6m & 1040s / 63s (16.5x) & 11.2gb / 0.5 gb (22.4x) \\ 
{\em S. aureus} single-cell & 58m / 0.3m & 5352s / 35s (153x) & 54.4gb / 0.4gb (136x) \\
{\em Deltaproteobacteria} single-cell & 67m / 0.4m & 4749s / 26s (182.7x) & 52.7gb / 0.4gb (131.8x) \\

\end{tabular}
\begin{flushleft}
\end{flushleft}
\label{tab:dngenome}
\end{table}


\begin{table}[!ht]
\caption{
\bf{Single-pass digital normalization to C=20 reduces computational
requirements for transcriptome assembly.}}


\begin{tabular}{|l|c|c|c|c|}

Data set & N reads pre/post & Assembly time pre/post & Assembly memory pre/post \\
 \hline \\
Yeast (Oases) & 100m / 9.3m & 181 min / 12 min (15.1x) & 45.2gb / 8.9gb (5.1x) \\
Yeast (Trinity) & 100m / 9.3m & 887 min / 145 min (6.1x) & 31.8gb / 10.4gb (3.1x) \\
Mouse (Oases) & 100m / 26.4m & 761 min/ 73 min (10.4x) & 116.0gb / 34.6gb (3.4x) \\
Mouse (Trinity) & 100m / 26.4m & 2297 min / 634 min (3.6x) & 42.1gb / 36.4gb (1.2x) \\
\end{tabular}

\begin{flushleft}
\end{flushleft}
\label{tab:dntrans}
\end{table}


\begin{table}[!ht]
\caption{
\bf{Digital normalization has assembler-specific effects on transcriptome
assembly.}}


\begin{tabular}{|l|c|c|c|c|}

Data set & Contigs $>$ 300 & Total bp $>$ 300 & Contigs $>$ 1000 & Total bp $>$ 1000 \\
\hline \\
Yeast (Oases) & 12,654 / 9,547 & 33.2mb / 27.7mb & 9,156 / 7,345 & 31.2mb / 26.4mb \\
Yeast (Trinity) & 10,344 / 12,092 & 16.2mb / 16.5mb & 5,765 / 6,053 & 13.6 mb / 13.1mb \\
Mouse (Oases) & 57,066 / 49,356 & 98.1mb / 84.9mb & 31,858 / 27,318 & 83.7mb / 72.4mb \\
Mouse (Trinity) & 50,801 / 61,242 & 79.6 mb / 78.8mb & 23,760 / 24,994 & 65.7mb / 59.4mb \\

\end{tabular}

\begin{flushleft}
\end{flushleft}
\label{tab:dntrans0}
\end{table}

\end{document}